\documentclass[preprint,showpacs,epsfig,eqsecnum,aps]{revtex4}
\usepackage{graphicx}
\usepackage{epsfig}
\begin{document}

\title{Double beta decay to the first $2^+$ state within a boson expansion formalism with a projected spherical
single particle basis}

\author{A. A. Raduta$^{a),b)}$, C.M. Raduta $^{b)}$}

\address{
$^{b)}$Department of Theoretical Physics and Mathematics,Bucharest University, POBox MG11,
Romania}

\address{$^{a)}$Institute of Physics and Nuclear Engineering, Bucharest, POBox MG6, Romania}

\begin{abstract}
The Gamow-Teller transition operator is written as a polynomial in the dipole proton-neutron and quadrupole charge conserving
QRPA boson operators, using the prescription of the boson expansion technique of Belyaev-Zelevinski type. Then, the $2\nu\beta\beta$ process ending on the first $2^+$ state in the daughter nucleus is allowed via  one, two and  three boson  states describing the odd-odd intermediate nucleus. The approach uses a single particle basis which is obtained by projecting out the good angular momentum from an orthogonal set of deformed functions. The basis for mother and daughter nuclei have different deformations. The GT transition amplitude as well as the half lives were calculated for ten transitions. Results are compared with the available data as well as with some predictions obtained with other methods.
\end{abstract}
\maketitle

One of the most exciting subject of nuclear physics  is that of double beta decay.
The interest is generated by the fact that in order to describe quantitatively
the decay rate one has to treat consistently the neutrino properties as well as the
nuclear structure features.
The process may take place in two distinct ways: a) by a $2\nu\beta\beta$ where
decay the initial nuclear system, the mother nucleus, is transformed in
  the final stable nuclear system, usually called
the daughter nucleus, two electrons and two anti-neutrinos;
b) by the $0\nu\beta\beta$ process where
the final state  does not involve any neutrino. The latter decay
mode is
especially interesting since one hopes that its discovery might provide a definite
answer to the question whether the neutrino is a Majorana or a Dirac particle.
 The contributions over
several decades have
been reviewed by many authors. 
\cite{PriRo,HaSt,Fass,Suh,Ver,Rad1}.

Although none of the double beta emitters is a
spherical nucleus
most formalisms use a single particle spherical basis.

In the middle of 90's we treated the $2\nu\beta\beta$
process in a pnQRPA formalism using a projected spherical single particle
basis which resulted in having a unified description of the process for
spherical and deformed nuclei \cite{Rad2,Rad3}. Recently the single particle basis \cite{Rad4,Rad44}
has been improved by accounting for the volume conservation while the mean
field is deformed \cite{Rad5,Rad6}. The improved basis has been used for
describing quantitatively the double beta decay rates as well as the corresponding half lives \cite{Rad55,Rad66}.
The results were compared with the available data as well as with the predictions of other formalisms.
The manners in which the physical observable is  influenced by the nuclear deformations of mother and daughter nuclei are in detail commented.
Two features of the deformed basis are essential: a) the single particle energy levels do not exhibit any gap; b) the pairing properties
of the deformed system are different from those of spherical system. These two aspects of the deformed nuclei affect the overlap matrix of the pnQRPA states of mother and daughter nuclei. Moreover, considering the Gamow-Teller (GT) transition operator in the single particle-space generated by the deformed mean-field, one obtains an inherent renormalization with respect to the one acting in a spherical basis.

In Ref. (\cite{Rad7}) we studied the higher pnQRPA effects on the GT transition amplitude, by means of the boson expansion technique for
 a spherical single particle basis. Considering higher order boson expansion terms in the transition operator, significant corrections
to the GT transition amplitude are obtained especially when the strength of the two body particle-particle  ($pp$) interaction   
approaches its critical value where the
lowest dipole energy is vanishing. As we showed in the quoted reference, there are transitions which are forbidden at the pnQRPA level but  allowed once the higher pnQRPA corrections are included.
An example of this type is the $2\nu\beta\beta$ decay leaving the daughter nucleus in a collective excited state $2^+$.
The electrons  resulting in this process can be distinguished from the ones associated to the ground to ground  transition by measuring, in coincidence, the gamma rays due to the transition $2^+\to 0^+$ in the daughter nucleus \cite{Baraba}.

The aim of this letter is to study the double beta decay $0^+\to 2^+$ where $0^+$ is the ground state of the emitter while $2^+$ is a single quadrupole phonon state describing the daughter nucleus. The procedure is the boson expansion method as formulated in our previous paper
\cite{Rad7} but using a projected spherical single particle basis.

In order to fix the necessary notations and to be self-contained, in the
present work we  describe briefly
the main ideas underlying the construction of the projected single
particle basis. 

The single particle mean field is determined by a particle-core Hamiltonian: 
\begin{equation}
\tilde{H}=H_{sm}+H_{core}-
M\omega_0^2r^2\sum_{\lambda=0,2}\sum_{-\lambda\le\mu\le\lambda}
\alpha_{\lambda\mu}^*Y_{\lambda\mu},
\label{hpc}
\end{equation}
where  $H_{sm}$ denotes the spherical shell model Hamiltonian while $H_{core}$ is
a harmonic quadrupole boson ($b^+_\mu$) Hamiltonian associated to a phenomenological core. 
The interaction of the two subsystems is accounted for by 
the third term of the above equation, written in terms of the shape coordinates $\alpha_{00}, \alpha_{2\mu}$.
The quadrupole shape coordinates are related to the quadrupole boson
 operators by the canonical transformation:
\begin{equation}
\alpha_{2\mu}=\frac{1}{k\sqrt{2}}(b^{\dagger}_{2\mu}+(-)^{\mu}b_{2,-\mu}),
\label{alpha2}
\end{equation}
where $k$ is an arbitrary C number. The monopole shape coordinate is   
determined from the volume conservation condition. In the quantized form, the result is:
\begin{equation}
\alpha_{00}=\frac{1}{2k^2\sqrt{\pi}}\left[5+\sum_{\mu}(2b^{\dagger}_{\mu}b_\mu
+(b^{\dagger}_{\mu}b^{\dagger}_{-\mu}+b_{-\mu}b_{\mu})(-)^{\mu})\right].
\label{alpha0}
\end{equation}
Averaging $\tilde{H}$ on the eigenstates of $H_{sm}$, hereafter denoted by
$|nljm\rangle$, one obtains a deformed boson Hamiltonian whose ground state 
is, in the harmonic limit, described by a coherent state
\begin{equation}
{\Psi}_g=exp[d(b_{20}^+-b_{20})]|0\rangle_b,
\label{psig}
\end{equation}
with $|0\rangle_b$ standing for the vacuum state of the boson operators and $d$ a real parameter 
which simulates the nuclear deformation.
On the other hand, the average of $\tilde{H}$ on ${\Psi}_g$ is  similar to
the Nilsson Hamiltonian \cite{Nils}.
Due to these properties, it is expected that the best trial functions 
to generate a spherical basis are:
\begin{equation}
{\Psi}^{pc}_{nlj}=|nljm\rangle{\Psi}_g.
\label{psipc}
\end{equation}
The projected states are obtained  by acting
on these deformed states with the projection operator
\begin{equation}
P_{MK}^I=\frac{2I+1}{8\pi^2}\int{D_{MK}^I}^*(\Omega)\hat{R}(\Omega)
d\Omega .
\label{pjmk}
\end{equation}
The subset of projected states :
\begin{equation}
\Phi_{nlj}^{IM}(d)={\cal N}_{nlj}^IP_{MI}^I[|nljI\rangle\Psi_g]\equiv
{\cal N}_{nlj}^I\Psi_{nlj}^{IM}(d) ,
\label{phiim}
\end{equation}
are orthogonal with the normalization factor denoted by ${\cal N}_{nlj}^I$.

 Although the projected states are associated to the particle-core system,
they can be used as a single particle basis. Indeed, when a matrix element of 
a particle like operator is calculated, 
the integration on the core collective coordinates is performed first,  which 
results in obtaining a final 
factorized expression: one factor carries the dependence on deformation and
one is a spherical shell model matrix element.

The single particle energies are approximated by the  average of the particle-core Hamiltonian $H'=\tilde{H}-H_{core}$
on the projected spherical states defined by Eq.(\ref{phiim}):
\begin{equation}
\epsilon_{nlj}^I=\langle\Phi_{nlj}^{IM}(d)|H'|\Phi_{nlj}^{IM}(d)\rangle .
\label{epsI}
\end{equation}
The off-diagonal matrix elements of $H'$ is ignored at this level. Their contribution is however considered when the
residual interaction is studied. 

As shown in Ref.\cite{Rad4}, the dependence of the new single particle energies on
deformation is similar to that shown by
the Nilsson model \cite{Nils}.
 The 
quantum numbers in the two schemes 
are however different. Indeed, here we generate from each j a multiplet of $(2j+1)$
states distinguished by the quantum number I, which plays the role of the Nilsson quantum number $\Omega$ and runs from 1/2 to j and moreover the energies
corresponding to the quantum numbers K
and -K are equal to each other.
On the other hand, for a given I there are $2I+1$ degenerate sub-states while the Nilsson states are only double degenerate. 
As explained in Ref.\cite{Rad4}, the redundancy
problem can be solved by changing the normalization of the model functions:
\begin{equation}
\langle\Phi_{\alpha}^{I M}|\Phi_{\alpha}^{I M}\rangle=1 \Longrightarrow \sum_{M}\langle\Phi_{\alpha}^{IM}|\Phi_{\alpha}^{IM}\rangle=2.
\label{newnorm}
\end{equation}
Due to this weighting factor the particle density function is providing the
consistency result that the number of particles which can be
distributed on the (2I+1) sub-states is at most 2, which agrees with the
Nilsson model.
Here $\alpha$ stands for the set of shell model quantum numbers $nlj$.
Due to this normalization, the states $\Phi^{IM}_{\alpha}$ used to calculate the
matrix elements of a given operator should be multiplied with the weighting factor $\sqrt{2/(2I+1)}$.

Finally, we recall a
fundamental result, obtained in Ref.\cite{Rad6}, concerning the product of two projected
 states, which comprises a product of two core components.
Therein we have proved that the
matrix elements of a two body interaction corresponding to the present scheme are
very close to the matrix elements corresponding to spherical
states projected from  a  deformed  state consisting of
two spherical single
  particle states times a single collective
  core wave function. The small discrepancies of the two types of matrix elements
  could be washed out by using  slightly different strengths for the two body interaction in the two
  methods.

As we already stated, in the present work we are interested to describe
the Gamow-Teller two neutrino double beta decay of an even-even deformed nucleus.
In our treatment the Fermi transitions, contributing about 20\% to the total rate, and the
``forbidden''
transitions are ignored, which is a reasonable approximation for
the two neutrino double beta decay in medium and heavy nuclei.
 The $2\nu\beta\beta$ process is conceived as two successive
single
$\beta^-$ transitions. The first transition connects the ground state of
the mother nucleus to a magnetic dipole state $1^+$ of the intermediate odd-odd nucleus which
subsequently decays to the first state $2^+$ of the daughter nucleus.
The second leg of the transition is forbidden within the pnQRPA approach but non-vanishing within a higher pnQRPA approach
 \cite{Rad7}. 
The states, involved in the $2\nu\beta\beta$
 process are described by the following many body Hamiltonian:
\begin{eqnarray}
H=&&\sum\ \frac{2}{2I+1}(\epsilon_{\tau\alpha I}-
\lambda_{\tau\alpha})c^{\dagger}_{\tau\alpha IM}c_{\tau \alpha IM}-\sum\frac{G_{\tau}}{4}
P^{\dagger}_{\tau \alpha I}P_{\tau\alpha I'}\nonumber \\
 &+& 2\chi\sum\beta^-_{\mu}(pn)\beta^+_{-\mu}(p'n')(-)^{\mu}
 -2\chi_1\sum P^-_{1\mu}(pn)P^+_{1,-\mu}(p'n')(-)^{\mu}\nonumber\\
&-&\sum_{\tau,\tau^{\prime}=p,n}X_{\tau,\tau^{\prime}}Q_{\tau}Q_{\tau^{\prime}}^{\dagger}.
\label{Has}
\end{eqnarray}
The operator $c^{\dagger}_{\tau\alpha IM}(c_{\tau\alpha IM})$
creates (annihilates) a particle of type $\tau$ (=p,n)
in the state $\Phi^{IM}_{\alpha}$, when acting on the vacuum
state $|0\rangle$. In order to simplify the notations, hereafter the set of
quantum numbers $\alpha(=nlj)$ will be omitted. The two body interaction
consists of three terms, the pairing, the dipole-dipole particle hole (ph) and
the particle-particle (pp) interactions. The corresponding strengths are
denoted by $G_{\tau},\chi,\chi_1$, respectively. All of them are separable
interactions, with the factors defined by the following expressions:
\begin{eqnarray}
P^{\dagger}_{\tau I}&=&\sum_{M}\frac{2}{2I+1}c^{\dagger}_{\tau IM}
c^{\dagger}_{\widetilde{\tau IM}},\nonumber\\
\beta^-_{\mu}(pn)&=&\sum_{M,M'}\frac{\sqrt{2}}{{\hat I}}
\langle pIM|\sigma_{\mu}|n I'M'\rangle \frac{\sqrt{2}}{{\hat {I'}}}
c^{\dagger}_{pIM}c_{nI'M'},\nonumber\\
P^-_{1\mu}(pn)& = & \sum_{M,M'} \frac{\sqrt{2}}{{\hat I}}\langle pIM|\sigma_{\mu}|nI'M'\rangle \frac{\sqrt{2}}{{\hat {I'}}}
c^{\dagger}_{pIM}c^{\dagger}_{\widetilde{nI'M'}},\nonumber\\
Q^{(\tau)}_{2\mu}&=&\sum_{i,k}q^{(\tau )}_{ik}\left(c^{\dagger}_ic_k\right)_{2\mu},\;q^{(\tau )}_{ik}=\sqrt{\frac{2}{2I_k+1}}\langle I_i||r^2Y_2||I_k\rangle.
\label{Psibeta}
\end{eqnarray}
The remaining operators from Eq.(\ref{Has}) can be obtained from the above defined operators
by hermitian conjugation.

The one body term and the pairing interaction terms are treated first through
the standard BCS formalism and consequently replaced by the quasiparticle
one body term $\sum_{\tau IM}E_{\tau}a^{\dagger}_{\tau IM}a_{\tau IM}$.
In terms of quasiparticle creation ($a^{\dagger}_{\tau IM}$) and annihilation
($a_{\tau IM}$) operators, related to the particle operators by means of the
Bogoliubov-Valatin transformation, the two body interaction terms, involved
in the model Hamiltonian, can be expressed just by replacing the
operators (3.2) by their quasiparticle images. Thus, the Hamiltonian terms describing the quasiparticle correlations become a quadratic 
expression in the dipole and quadrupole two quasiparticles and quasiparticle density operators:
\begin{eqnarray}
A^{\dagger}_{1\mu}(pn)&=&\sum_{m_p,m_n}C^{I_p\; I_n \; 1}_{m_p \; m_n \; \mu}
a^{\dagger}_{pI_pm_p}a^{\dagger}_{n I_n m_n},\nonumber\\
B^{\dagger}_{1\mu}(pn)&=&\sum_{m_p,m_n}C^{I_p\; I_n \; 1}_{m_p \;-m_n \; \mu}
a^{\dagger}_{pI_pm_p}a_{n I_n m_n}(-)^{I_n-m_n},\nonumber\\
A^{\dagger}_{2\mu}(\tau \tau^{\prime})&=&\sum_{m_{\tau},m_{\tau^{\prime}}}C^{I_{\tau}\; I_{\tau^{\prime}} \; 1}_{m_{\tau} \; 
m_{\tau^{\prime}} \; \mu}a^{\dagger}_{\tau I_{\tau}m_{\tau}}a^{\dagger}_{\tau^{\prime} I_{\tau^{\prime}} m_{\tau^{\prime}}},\nonumber\\
B^{\dagger}_{2\mu}(\tau \tau^{\prime})&=&\sum_{m_{\tau},m_{\tau^{\prime}}}C^{I_{\tau}\; I_{\tau^{\prime}} \; 2}_{m_{\tau} \;
-m_{\tau^{\prime}} \; \mu}
a^{\dagger}_{\tau I_{\tau} m_{\tau}}a_{\tau^{\prime} I_{\tau^{\prime}} m_{\tau^{\prime}}}(-)^{I_{\tau^{\prime}}-m_{\tau^{\prime}}}
.
\label{A1B1}
\end{eqnarray}

Since the $pnQRPA$ treatment of the dipole-dipole interaction in the particle-hole ($ph$) and $pp$ channels
run in an identical way as in our previous publications \cite{Rad55,Rad66}, here we don't give any detail about building the dipole proton-neutron phonon operator
 :
\begin{equation}
\Gamma^{\dagger}_{1\mu}=\sum_{k}[X_1(k)A^{\dagger}_{1\mu}(k)-Y_1(k)A_{1,-\mu}(k)(-)^{1-\mu}],
\label{Gama}
\end{equation}
We just mention that the amplitude are determined by the pnQRPA equations and the normalization condition.

The charge conserving $QRPA$ bosons 
\begin{equation}
\Gamma^{\dagger}_{2\mu}=\sum_{k}[X_2(k)A^{\dagger}_{2\mu}(k)-Y_1(k)A_{2,-\mu}(k)(-)^{\mu}],\;k=(p,p^{\prime}),(n,n^{\prime}),
\label{Gama2}
\end{equation}
are determined by the  $QRPA$ equations associated to the matrices:
\begin{eqnarray}
{\cal A}_{\tau\tau^{\prime}}(ik;i^{\prime}k^{\prime})&=&\delta_{\tau\tau^{\prime}}\delta_{ii^{\prime}}\delta_{kk^{\prime}}(E^{\tau}_i+E^{\tau}_k)-X_{\tau\tau^{\prime}}\left(q^{(\tau)}_{ik}\xi^{(\tau)}_{ik}\right)\left(q^{(\tau)}_{i^{\prime}k^{\prime}}\xi^{(\tau)}_{i^{\prime}k^{\prime}}\right),\nonumber\\
{\cal B}_{\tau\tau^{\prime}}(ik;i^{\prime}k^{\prime})&=&
-X_{\tau\tau^{\prime}}\left(q^{(\tau)}_{ik}\xi^{(\tau)}_{ik}\right)\left(q^{(\tau)}_{i^{\prime}k^{\prime}}\xi^{(\tau)}_{i^{\prime}k^{\prime}}\right),\;i\leq k,\;i^{\prime}\leq k^{\prime}.
\end{eqnarray}
where 
\begin{equation}
\xi^{(\tau)}_{ik}=\frac{\left(U^{\tau}_iV^{\tau}_k+U^{\tau}_kV^{\tau}_i\right)}{\sqrt{1+\delta_{i,k}}}.
\end{equation}
 In order to distinguish between the phonon operators acting in the RPA space associated to the mother and daughter nuclei respectively, one needs an additional index. Also, an index labeling the solutions of the RPA equations is necessary.  Thus, the two kinds of bosons will be denoted by:
\begin{equation}
\Gamma^{\dagger}_{1\mu}(jk),\;j=i,f;\; k=1,2,...N^{(1)}_s;\Gamma^{\dagger}_{2\mu}(jk),\;j=i,f;\; k=1,2,...N^{(2)}_s.
\label{rpast}
\end{equation}
Acting with $\Gamma^{\dagger}_{1\mu}(ik)$ and $\Gamma^{\dagger}_{1\mu}(fk)$ on the vacuum states $|0\rangle_i$ and $|0\rangle_f$ respectively, one obtains two sets of non-orthogonal states describing the intermediate odd-odd nucleus. By contrast, the states
$\Gamma^{\dagger}_2(ik)|0\rangle_i$ and $\Gamma^{\dagger}_2(fk)|0\rangle_f$ describe different nuclei, namely the initial and final ones 
participating in the process of $2\nu\beta\beta$ decay. The mentioned indices are however  omitted whenever their presence is not necessary.

Within the boson expansion formalism the transition GT operators are written as polynomial expansion in terms of the QRPA boson operators with the expansion coefficients determined such that the mutual commutation relations of the constituent operators $A^{\dagger}_{1\mu}(pn), A_{1\mu}(pn), 
B^{\dagger}_{1\mu}(pn), B_{1\mu}(pn)$ be preserved in each order of approximation \cite{BeZe}. One arrives at the expressions:   
\begin{eqnarray}
& &A^{\dagger}_{1\mu}(j_pj_n)=\sum_{k_1}\left\{{\cal A}^{(1,0)}_{k_1}(j_pj_n)\Gamma^{\dagger}_{1\mu}(k_1)+
{\cal A}^{(0,1)}_{k_1}(j_pj_n)\Gamma_{1-\mu}(k_1)(-)^{1-\mu}\right\}\nonumber\\
&+&\sum_{k_1,k_2,k_3;l=0,2}
\left\{{\cal A}^{(3,0);l}_{K_3k_2k_1}(j_pj_n)\left[\left(\Gamma^{\dagger}_2(k_3)\Gamma^{\dagger}_2(k_2)\right)_l\Gamma^{\dagger}_1(k_1)\right]_{1\mu}
+
{\cal A}^{(0,3);l}_{K_3k_2k_1}(j_pj_n)\left[\left(\Gamma_2(k_3)\Gamma_2(k_2)
\right)_l\Gamma_1(k_1)\right]_{1\mu}\right\}\nonumber\\
&+&
\sum_{k_1,k_2,k_3;l=0,2}
\left\{{\cal A}^{1;(2\bar{2})l}_{K_1k_2k_3}(j_pj_n)\left[\Gamma^{\dagger}_1(k_1)\left(\Gamma^{\dagger}_2(k_2)\Gamma_2(k_3)
\right)_l\right]_{1\mu}
+
{\cal A}^{(2\bar{2})l;1}_{K_3k_2k_1}(j_pj_n)\left[\left(\Gamma^{\dagger}_2(k_3)\Gamma_2(k_2)\right)_l\Gamma_1(k_1)
\right]_{1\mu}\right\}\nonumber\\
&&B^{\dagger}_{1\mu}(j_pj_n)=\sum_{k_1k_2}\left\{
{\cal B}^{(2,0)}_{k_1k_2}(j_pj_n)
\left[\Gamma^{\dagger}_1(k_1)\Gamma^{\dagger}_2(k_2)\right]_{l\mu}+
{\cal B}^{(0,2)}_{k_1k_2}(j_pj_n)
\left[\Gamma_1(k_1)\Gamma_2(k_2)\right]_{l\mu}\right . \nonumber\\
&+&{\cal B}^{11;12}_{k_1k_2}(j_pj_n)
\left .\left[\Gamma^{\dagger}_1(k_1)\Gamma_2(k_2)\right]_{l\mu}+
{\cal B}^{11;2l}_{k_1k_2}(j_pj_n)
\left[\Gamma^{\dagger}_1(k_2)\Gamma_1(k_1)\right]_{l\mu}\right \},
\label{bosexp}
\end{eqnarray}
where the expansion coefficients are those given in Ref.\cite{Rad7} .
If the energy carried by leptons in the intermediate state is approximated by
the sum of the rest energy of the emitted electron and half the Q-value of
the double beta decay process
\begin{equation}
\Delta E=m_ec^2+\frac{1}{2}Q^{(0\to 2)}_{\beta\beta},
\label{DeltaE}
\end{equation} 
the reciprocal value of the $2\nu\beta\beta$ half life can be factorized as:
\begin{equation}
T^{2\nu}_{1/2}(0^+_i\to 2^+_f)^{-1}=F_2|M^{(02)}_{GT}|^2,
\label{hlife}
\end{equation}
where $F_2$ is the Fermi integral which characterizes the phase space of the process while the second factor is the GT transition amplitude which, in the second order of perturbation theory, has the expression:
\begin{equation}
M^{(02)}_{GT}=\sqrt{3}\sum_{k,m}\frac{_i\langle0 ||\beta^+||k,m\rangle _{i}~
_i\langle k,m|k^{\prime},m^{\prime}\rangle _f~_f\langle k^{\prime},m^{\prime}||\beta^+||2^+_1\rangle _f}
{(E_{k,m}+\Delta E_2)^3}.
\label{M02}
\end{equation}
Here $\Delta E_2=\Delta E +E_{1^+}$, with $E_{1^+}$ standing for the experimental energy  for the first state $1^+$.
The intermediate states $|k,m\rangle$ are k-boson states with $k=1,2,3$ labeled by the index m, specifying the spin and the ordering label of the RPA roots. Inserting the boson expansions
from Eq.(\ref{bosexp}) into the expression of the $\beta^+$ transition operator one can check that the following non-vanishing factors, at numerator,
show up:
\begin{eqnarray}
& &{_i}\langle 0||\Gamma_1(i,k_1)||1,1_{k_1}\rangle_i {_f}\langle 1,1_{k_2}||\Gamma^{\dagger}_1(f,k_2)\Gamma_2(f,1)||1,2_1\rangle_f, \nonumber\\
& &{_i}\langle 0||\Gamma_1(i,k_1)\Gamma_2(i,k_2)||2,1_{k_1}2_{k_2}\rangle _i {_f}\langle 2,1_{j_1}2_1||\Gamma^{\dagger}_1(f,j_1)||1,2_1\rangle _f ,\nonumber\\
& &{_i}\langle 0||\Gamma_1(i,k_1)\Gamma_2(i,k_2)\Gamma_2(i,k_3)||3,1_{k_1}2_{k_2}2_{k_3}\rangle_i {_f}\langle 3,1_{j_1}2_{j_2}2_1||
\Gamma^{\dagger}_1(f,j_1)\Gamma^{\dagger}_2(f,j_2)||1,2_1\rangle _f ,\nonumber\\
& &{_i}\langle 0||\Gamma_1(i,k_1)\Gamma_2(i,k_2)||2,1_{k_1}2_{k_2}\rangle_i {_f}\langle 2,1_{j_1}2_{j_2}||
\Gamma^{\dagger}_1(f,j_1)\Gamma^{\dagger}_2(f,j_2)\Gamma_2(f,1)||1,2_1\rangle _f.
\label{02mel}
\end{eqnarray}
 The term $E_{k,m}$ from the denominator of
 Eq. (\ref{M02}) is the average of the energies of the mother and daughter states $|k,m\rangle$ normalized to the average energy of the first $pnQRPA$ states $1^+$ in the initial and final nuclei.
 Calculations were performed for the following 10 double beta emitters:$^{48}$Ca,$^{96}$Zr, $^{100}$Mo, $^{104}$Ru, $^{110}$Pd,
$^{116}$Cd, $^{128}$Te,$^{130}$Te, $^{134}$Xe, $^{136}$Xe.
Since  the single particle space,
the pairing interaction treatment, and the pnQRPA description of the dipole  states describing the intermediate odd-odd nuclei used in the present paper are identical with those from Refs.\cite{Rad55,Rad66} for ground to ground transition, we don't present them again. The strength of the $QQ$ interaction was fixed by requiring that the first root  of the QRPA equation for the quadrupole charge conserving boson is close to the experimental energy of the first $2^+$ state. The results of the fitting procedure are given in Table I.

Having the RPA states defined, the GT amplitude has been calculated by means of Eq.(\ref{M02}), while the half life with Eq.(\ref{hlife}).
The Fermi integral for the transition $0^+\to 2^+$ was computed by using the analytical result given in Ref.(\cite{Suh}).

\begin{table}[ht!]
\begin{tabular}{|c|c|c|c|}
\hline
Nucleus& $E^{exp.}_{2^+}$ [keV]  &$E^{th.}_{2^+}$ [keV] & $b^4X_{pp}[keV]$\\
       \hline
$^{48}$Ca& 983 & 983  & 0.0713  \\
$^{48}$Ti& 983.52 & 979.02  & 42.8  \\
$^{76}$Ge& 562.93 & 558.88  & 50.8\\
$^{76}$Se& 559.10 & 558.87  & 65.2\\
$^{96}$Zr& 1750.49 & 1465.6  & 2.\\
$^{96}$Mo& 778.24 & 776.8  & 38.1\\
$^{100}$Mo& 535.57 & 534.4 &31.5 \\
$^{100}$Ru& 539.5 & 536.1 & 19.7 \\
$^{104}$Ru& 358.03 & 358.45 & 29.8  \\
$^{104}$Pd& 555.81 & 561.83 & 20.9  \\
$^{110}$Pd& 373.8& 370.45 &44.65 \\
$^{110}$Cd& 657.76& 662.8 & 25.1 \\
$^{116}$Cd& 513.49 & 514.5 & 30.5 \\
$^{116}$Sn& 1293.56 &1179.16 & 7.0 \\
$^{128}$Te& 743.22 & 746.12 & 12.12 \\
$^{128}$Xe& 442.91 & 449.58 & 19.43 \\
$^{130}$Te& 839.49 & 831.03 & 12.12 \\
$^{130}$Xe& 536.07 & 534.2 & 17.28 \\
$^{134}$Xe& 847.04 & 841.75 & 20.0  \\
$^{134}$Ba& 604.72 & 607.98 & 17.56  \\
$^{136}$Xe& 1313.027 & 1314.9 & 16.37 \\
$^{136}$Ba& 818.49 & 810.3 & 14.82 \\
\hline
\end{tabular}
\caption{The experimental and calculated energies for the first $2^+$ states in mother and daughter nuclei are given. The strength parameter of the quadrupole-quadrupole interaction was fixed so that the experimental energies are reproduced. In our calculations we considered
$X_{pp}=X_{nn}=X_{pn}$. The oscillator length is denoted by $b=(\hbar/M\omega)^{1/2}$.
 }
\end{table}

\begin{table}[h!]
\begin{tabular}{|c|c|c|c|c|c|c|}
\hline
Nucleus&$Q^{2^+}_{\beta\beta}$&$\Delta E_2 [MeV]$&$|M^{(0\to 2)}_{GT}|[MeV^{-1}]$&
\multicolumn{3}{c}{$T^{(0\to 2)}_{1/2}[yr]$}\\ \cline{5-7}
       &units of $m_ec^2$&    &     &  present & Exp. & Suhonen \cite{Suh1}\\
       \hline
$^{48}$Ca& 6.432 & 2.473  & 0.901$\cdot 10^{-3}$&1.72$\cdot10^{24}$ &  &  \\
$^{76}$Ge& 2.894 & 1.295  & 0.558$\cdot 10^{-3}$&5.75$\cdot10^{28}$ &
$> $1.1$\cdot 10^{21}$  &1.0$\cdot 10^{26}$  \\
$^{96}$Zr& 5.033 & 2.913  & 0.834$\cdot 10^{-3}$&2.27$\cdot10^{25}$ &
$> $ 7.9$\cdot 10^{19}$  & 4.8$\cdot 10^{21}$ \\
$^{100}$Mo& 4.874 & 1.756 & 0.136$\cdot 10^{-2}$&1.21$\cdot10^{25}$ &
$> $1.6$\cdot 10^{21} $ &3.9$\cdot 10^{24}$  \\
     &   &   &    &   &    &$^{a)}$2.5$\cdot 10^{25}$\\
     &   &   &    &   &    &$^{b)}$1.2$ \cdot 10^{26}$ \\
$^{104}$Ru& 1.456 & 0.883 & 0.028 &6.2$\cdot10^{28}$ &  &  \\
$^{110}$Pd& 2.646 & 1.182 & 0.050 &1.48$\cdot10^{25}$ &  &  \\
$^{116}$Cd& 2.967 & 1.269 & 0.507$\cdot 10^{-2}$&3.4$\cdot10^{26}$ &
$> $2.3$\cdot 10^{21}$  & 1.1$\cdot 10^{24}$ \\
$^{128}$Te& 0.836 & 1.305 & 0.229$\cdot 10^{-2}$&4.7$\cdot10^{33}$ &
$> $4.7$\cdot 10^{21}$  &1.6$\cdot 10^{30}$  \\
$^{130}$Te& 3.902 & 2.358 & 0.620$\cdot 10^{-3}$&6.94$\cdot10^{26}$ &
>4.5$\cdot 10^{21}$  & 2.7$\cdot 10^{23}$ \\
$^{134}$Xe& 0.460 & 0.806 & 0.621$\cdot 10^{-2}$&5.29$\cdot10^{35}$ &  &  \\
$^{136}$Xe& 3.251 & 1.518 & 0.249$\cdot 10^{-2}$&3.88$\cdot10^{26}$ &  &
2.0$\cdot 10^{24}$ \\
\hline
\end{tabular}
\caption{The results for the GT transition amplitudes as well as for the half lives of the double beta decay
  $0^+\to 2^+$ are given. Also the $Q$ values of the transitions are given in units of
 $m_ec^2$. $\Delta E_2$ is the energy shift defined in the text.
For comparison, we give also the available experimental results as well as some theoretical predictions obtained with other formalisms.
For $^{100}$Mo we mention the result of Ref. (\cite{Hir}) obtained with an SU(3) deformed single particle basis $^{a)}$ and with a spherical basis
 $^{b)}$.}
\end{table}

The final results are collected in Table II. Therein one may find also the available experimental data as well as some theoretical results obtained with other approaches. One notices that the half life is influenced  by both the phase space integral (through the Q-value) and 
 the single particle properties which determine the transition amplitude. Indeed, for $^{128}$Te and $^{134}$Xe the small Q-value causes a very large half life, while in $^{48}$Ca the opposite situation is met. By contrary the Q value of $^{110}$Pd is about the same as for 
$^{76}$Ge but, due to the specific single particle and pairing properties of the orbits participating coherently to the process, the half life for the former case  is more than three orders of magnitude less than in the later  situation. Since in Ref.\cite{Suh1} a spherical single particle basis and a renormalized pnQRPA approach are used, the transition matrix elements are larger  and  the half lives shorter than in our 
calculations, although the higher RPA formalism in the quoted reference is similar \cite{Rad15} to the one proposed by us in Ref.\cite{Rad7} and used in the present work.
It is worth mentioning the good agreement between our prediction for $^{100}$Mo and that of Ref.\cite{Hir} obtained with a deformed SU(3) single particle basis. 

It is worth mentioning that the double beta transitions to excited states have been considered by several authors in the past, but the calculations emphasized the role of the transition operator and some specific selection rules. Many of calculations regarded the neutrinoless process. Thus, in Ref.\cite{Verg2} it was shown that 
the neutrinoless  transition to the excited $0^+$ for medium heavy nuclei might be  characterized by matrix element which are larger than 
that of ground to ground transition  and that happens since in the first transition, the change of the $K$ quantum number is less.  
In Ref.\cite{Rosen} it has been stated that the $0^+\to 2^+$ matrix element depends on the left-right current coupling and not on the neutrino mass.
However according to the calculations of Haxton {\it et al}\cite{HaSt}, the matrix element is suppressed and therefore it is not possible to extract a meaningful parameter for the left-right coupling. Although the transition operator might have a complex structure, many calculations have been performed with the approximate interaction $[\sigma(1)\times\sigma(2)]^{\lambda=2}t_+(1)t_+(2)$ in order to test some selection rules. Thus,  this interaction was used in Ref.\cite{Za} for the transition $0^+\to 2^+$ of $^{48}$Ca, using a single $j$ calculation. It has been proved that the matrix element for this transition is suppressed due to  the signature selection rules.
The transition to $0^+_1$ was examined for $A=76,82,100,136$ nuclei by assuming light and heavy Majorana neutrino exchange mechanism and triliniar R-parity contribution. Higher RPA  as well as renormalization effects for the nuclear matrix elements were included \cite{Rad20}.

Here we show that the transition $0^+\to 2^+$ in a $2\nu\beta\beta$ process is allowed by renormalizing the GT transition operator with some higher RPA corrections which results in making the matrix elements from Eq.(\ref{02mel}) non-vanishing.

\end{document}